\documentclass[12pt]{iopart}
\def\0{\mbox{\tiny $0$}}
\def\1{\mbox{\tiny $1$}}
\def\2{\mbox{\tiny $2$}}
\def\3{\mbox{\tiny $3$}}
\def\4{\mbox{\tiny $4$}}
\def\5{\mbox{\tiny $5$}}
\def\6{\mbox{\tiny $6$}}
\def\7{\mbox{\tiny $7$}}
\def\8{\mbox{\tiny $8$}}
\def\9{\mbox{\tiny $9$}}

\def\n{\mbox{\tiny $n$}}
\def\k{\mbox{\tiny $k$}}
\def\kk{\mbox{\small $k$}}

\def\f14{\mbox{\tiny $\frac{1}{4}$}}

\def\L{\mbox{\tiny $L$}}
\def\R{\mbox{\tiny $R$}}
\def\B{\mbox{\tiny $B$}}
\def\ii{\mbox{\tiny $i$}}

\def\z{\mbox{\tiny $z$}}
\def\s{\mbox{\tiny $s$}}
\def\r{\mbox{\tiny $r$}}

\def\x{\mbox{\tiny $x$}}
\def\y{\mbox{\tiny $y$}}

\def\j{\mbox{\tiny $j$}}
\def\mi{\mbox{\tiny $-$}}
\def\ig{\mbox{\tiny $=$}}
\def\pl{\mbox{\tiny $+$}}
\def\ppm{\mbox{\tiny $\pm$}}

\def\bb#1{\mbox{\footnotesize $(#1)$}}

\begin{document}

\title[Chirality dynamics for a fermionic particle non-minimally coupling with {\boldmath$B$}]{Chirality dynamics for a fermionic particle non-minimally coupling with an external magnetic field}

\author{A. E. Bernardini\dag\ 
\footnote[3]{alexeb@ifi.unicamp.br}
}

\address{\dag\ Instituto de Física Gleb Wataghin, UNICAMP,\\
PO Box 6165, SP 13083-970, Campinas, SP, Brazil}

\date{\today}

\begin{abstract}
We proceed with the construction of normalizable Dirac wave packets for treating chiral oscillations in the presence of an external magnetic field.
Both chirality and helicity quantum numbers correspond to variables of fundamental importance in the study of chiral interactions, in particular, in the context of neutrino physics.
In order to clarify a subtle aspect in the confront of such concepts which, for massive particles, represent different physical quantities, we are specifically interested in quantifying chiral oscillations for a {\em fermionic} Dirac-{\em type} particle (neutrino) non-minimally coupling with an external magnetic field {\boldmath$B$} by solving the correspondent interacting Hamiltonian (Dirac) equation.
The viability of the intermediate wave packet treatment becomes clear when we assume {\boldmath$B$} orthogonal/parallel to the direction of the propagating particle.
\end{abstract}




\maketitle 

Since those early days when Dirac had derived the relativistic wave equation for a free propagating
electron \cite{Dir28}, several efforts have been produced in the literature to solve the Dirac equation
with other analytical forms of interacting potentials, from central potential solutions \cite{Esp99,Alh05}
to recent theoretical attempts to describe quark confinement \cite{Ein81,Bak95}.
In fact, obtaining exact solutions of a generic class of Dirac wave equations becomes important since,
for many times, the conceptual understanding of physics can only be brought about by such solutions.
These solutions also correspond to valuable means for checking and improving models and numerical methods
for solving complicated physical problems.
In the context in which we intend to explore the Dirac formalism,
we can take into account the effect of neutrino spin flipping attributed to some dynamic external \cite{Oli90}
interacting processes which come from the non-minimal coupling of a magnetic moment with an external electromagnetic
field \cite{Vol81} and which was formerly supposed to be a relevant effect in the context of the solar-neutrino puzzle
by suggesting an explanation for the LSND anomaly \cite{Ana98,Agu01,Ban03} \footnote{In fact, the advent of the neutrino physics \cite{Zub98,Sch03,Alb03}
has stirred up an increasing number of theoretical works where
the quantum oscillation mechanisms \cite{Beu03,Vog04,Giu98,Giu02,Ber05A} have been deeply analyzed.
Although clear experimental evidences are still missing, it is widely believed that neutrino mixing
and oscillations are the basic tool for further understanding of neutrino phenomenology.}.

In the correspondent theoretical framework, we can notice a more extended discussion involving the evolution equation of the covariant neutrino spin operator in the Heisenberg representation in presence of general external fields \cite{Stu02}.
Such an analysis involves both electromagnetic and weak interactions with background matter \cite{Stu02,Stu03}.
In a similar context, some developments considering the minimal coupling of an electrically charged particle with an external magnetic field were also performed for an electron described by the Dirac equation \cite{Ter90}
and the theory of spin light of neutrino in matter and electromagnetic fields has been extensively studied \cite{Stu04}.

We know, however, that independently of any external electromagnetic field, since neutrinos are detected essentially via V-A
charged weak currents, the chiral oscillation mechanism by itself
may even provide a solution to the posed anomaly.
The point is that an erroneous correspondence is supposed to cause some confusion in the literature where the concept of
helicity has been currently used in the place of chirality.
In the standard model of flavor-changing interactions, neutrinos with positive
chirality are decoupled from neutrino absorbing charged weak currents \cite{DeL98}.
Consequently, neutrinos with positive chirality sterile with respect to weak interactions.
By reporting to the formalism with Dirac wave packets \cite{Zub80,Ber04}, which leads to the
study of chiral oscillations \cite{DeL98}
(in vacuum), we are now interested in appointing the modifications
of the chirality dynamics which
are observed when the neutrino electrodynamics is accurately discussed.

In this context, our aim in this manuscript is to try to accommodate
the Dirac wave packet formalism \cite{Zub80,Ber04} and
a further class of static characteristics of neutrinos, namely, the (electro)magnetic
moment which appears in a Lagrangian with non-minimal coupling.
By following this line of reasoning, we are not only interested in solving a ``modified'' Dirac wave equation
but also in constructing Dirac wave packets which can be used for describing the dynamics of chirality,
a physical variable which is relevant in the context of the quantum oscillation phenomena.

Despite their electric charge neutrality, neutrinos can interact with a photon through loop (radiative) diagrams.
The Lagrangian for the interaction between a fermionic field $\psi\bb{x}$ and an electromagnetic
field written in terms of the field-strength tensor $F^{\mu\nu}\bb{x}= \partial^{\mu}A^{\nu}\bb{x} - \partial^{\nu}A^{\mu}\bb{x}$ is given by
\begin{equation}
\mathcal{L} = \frac{1}{2}\,\overline{\psi}\bb{x} \,\sigma_{\mu\nu}\left[\mu\, F^{\mu\nu}\bb{x} - d \,\mathcal{F}^{\mu\nu}\bb{x}\right] \psi\bb{x} + h.c.
\end{equation}
where $x = \bb{t, \mbox{\boldmath$x$}}$, $\sigma_{\mu\nu} = \frac{i}{2}[\gamma_{\mu},\gamma_{\nu}]$,
the {\em dual} field-strength tensor $\mathcal{F}^{\mu\nu}\bb{x}$ is given by $\mathcal{F}^{\mu\nu}\bb{x}= \frac{1}{2}\epsilon^{\mu\nu\lambda\delta} F^{\lambda\delta}\bb{x}$\footnote{$\epsilon^{\mu\nu\lambda\delta}$ is the totally fourth rank antisymmetric tensor.} and  
the coefficients $\mu$ and $d$ represent, respectively, the magnetic and the electric dipole moment which
establish the neutrino electromagnetic coupling.
One can notice that we have not discriminated the flavor/mass mixing elements in the above interacting Lagrangian since
we are indeed interested in the
physical observable dynamics ruled by the Hamiltonian 
\begin{eqnarray} 
\mathit{H} &=& \mbox{\boldmath$\alpha$}\cdot \mbox{\boldmath$p$} + \beta m 
		- \beta\left[\frac{\sigma_{\mu\nu}}{2}\left(\mu \,F^{\mu\nu}\bb{x}- d \mathcal{F}^{\mu\nu}\bb{x}\right) + h.c.\right]\nonumber\\
 		   &=& \mbox{\boldmath$\alpha$}\cdot \mbox{\boldmath$p$} + \beta \left[m - \mu\, \mbox{\boldmath$\Sigma$}\cdot \mbox{\boldmath$B$}\bb{x} + d \, \mbox{\boldmath$\Sigma$}\cdot \mbox{\boldmath$E$}\bb{x}\right] 
\label{01},~~
\end{eqnarray} 
where, in terms of the Dirac matrices, $\mbox{\boldmath$\alpha$} = \sum_{\k \ig \1}^{\3} \alpha_{\k}\hat{\kk} = \sum_{\k \ig \1}^{\3} \gamma_{\0}\gamma_{\k}\hat{\kk}$,
$\beta = \gamma_{\0}$, and $\mbox{\boldmath$B$}\bb{x}$ and $\mbox{\boldmath$E$}\bb{x}$ are respectively the magnetic and electric fields.
In fact, the Eq.~(\ref{01}) could be extended to an equivalent matrix representation with flavor and mass mixing elements 
where the diagonal (off-diagonal) elements described by $\mu_{\ii,\j}(m_{\ii,\j})$ and $d_{\ii,\j}(m_{\ii,\j})$,
where $i, \, j$ are mass indices, would be called diagonal (transition) moments.
In this context, for both Dirac and Majorana neutrinos, we could have transition amplitudes with non-vanishing 
magnetic and electric dipole moments \cite{Sch1,Mar1,Akh1}.
Otherwise, the CP invariance holds the diagonal electric dipole moments null \cite{Kim93}.
Specifically for Majorana neutrinos, 
it can be demonstrated that the diagonal magnetic and electric dipole moments vanish if $CPT$ invariance is assumed \cite{Sch1}.

Turning back to the simplifying example of diagonal moments and assuming CP and CPT invariance, 
we can restrict our analysis to the coupling with only an external magnetic field $\mbox{\boldmath$B$}\bb{x}$ by setting $d = 0$.
From this point, the expression for $\mu$ can be found from Feynman diagrams for magnetic moment corrections \cite{Kim93} and turns out to be
proportional to the neutrino mass (matrix),
\begin{equation} 
\mu = \frac{3\, e \,G}{8 \sqrt{2}\pi^{\2}} m = \frac{3\, m_e \,G}{4 \sqrt{2}\pi^{\2}}\, \mu_{\B} \, m_{\nu}
= 2.7 \times 10^{\mi \1\0}\,\mu_{\B}\,\frac{m_{\nu}}{m_N}
\end{equation} 
where $G$ is the Fermi constant and $m_{N}$ is the nucleon mass\footnote{We are using some results of the
standard $SU(2)_{L} \otimes U(1)_{Y}$ electroweak theory \cite{Gla61}.}.
In particular, for $m_{\nu}\approx 1 \, eV$, the magnetic moment introduced by the above formula is exceedingly small
to be detected or to affect astrophysical or physical processes.

Since we are interested in constructing the dynamics ruled by the Hamiltonian of Eq.~(\ref{01}),
we firstly observe that the free propagating momentum is not a conserved quantity,
\begin{equation} 
\frac{d~}{dt}\langle\mbox{\boldmath$p$}\rangle \,=\, i\langle\left[ \mathit{H} , \mbox{\boldmath$p$}\right]\rangle\,=\, Re\bb{\mu} \langle \beta\, \mbox{\boldmath$\nabla$} \left(\mbox{\boldmath$\Sigma$}\cdot \mbox{\boldmath$B$}\bb{x}\right)\rangle
\label{02}
\end{equation} 
In the same way, the particle velocity given by
\begin{equation} 
\frac{d~}{dt}\langle\mbox{\boldmath$x$}\rangle \,=\, i\langle\left[ \mathit{H} , \mbox{\boldmath$x$}\right]\rangle\,=\,\langle\mbox{\boldmath$\alpha$}\rangle 
\label{03}
\end{equation}
comes out as a non-null value.
After solving the $\mbox{\boldmath$x$}\bb{t}$($\mbox{\boldmath$\alpha$}\bb{t}$) differential equation,
it is possible to observe that, in addition to a uniform motion, the fermionic particle
executes very rapid oscillations known as {\em zitterbewegung} \cite{Sak87}.
By following an analogous procedure for the Dirac chiral operator $\gamma_{\5}$,
newly recurring to the equation of the motion, it is possible to have
the chirality and the helicity dynamics respectively given by
\begin{eqnarray} 
\frac{d~}{dt}\langle \gamma^5 \rangle &=& 2\,i \langle\gamma_{\0}\,\gamma_{\5} \left[m - \mu\, \mbox{\boldmath$\Sigma$}\cdot \mbox{\boldmath$B$}\bb{x}\right]\rangle
\label{04}
\end{eqnarray} 
and
\begin{eqnarray} 
\frac{d~}{dt}\langle h \rangle &=& \frac{1}{2}\,\mu\,\langle \gamma_{\0} \left[(\mbox{\boldmath$\Sigma$}\cdot\mbox{\boldmath$\nabla$})(\mbox{\boldmath$\Sigma$}\cdot\mbox{\boldmath$B$}\bb{x})
+ 2(\mbox{\boldmath$\Sigma$}\times \mbox{\boldmath$B$}\bb{x})\cdot \mbox{\boldmath$p$}\right]\rangle
\label{05}
\end{eqnarray} 
where we have alternatively defined the particle helicity as 
the projection of the spin angular momentum onto the vector momentum,
$h = \frac{1}{2}\mbox{\boldmath$\Sigma$}\cdot{\mbox{\boldmath$p$}}$ (with
${\mbox{\boldmath$p$}}$ in place of $\hat{\mbox{\boldmath$p$}}$).
From Eqs.~(\ref{04}-\ref{05}) we can state that if a neutrino has an intrinsic magnetic moment and passes through a region filled by
an magnetic field, the neutrino helicity can flip in a completely different way from how chiral oscillations
evolve in time.
In the non-interacting case, it is possible to verify that the time-dependent averaged value of
the Dirac chiral operator $\gamma_{\5}$ has an oscillating behavior \cite{DeL98} very similar to
the rapid oscillations of the position.
The Eqs.~(\ref{04}-\ref{05}) can be reduced to the non-interacting case by setting $\mbox{\boldmath$B$}\bb{x} = 0$ so that
\begin{equation} 
\frac{d~}{dt}\langle h \rangle \,=\, i\langle\left[ \mathit{H} , h\right]\rangle\,=\,- \langle(\mbox{\boldmath$\alpha$} \times \mbox{\boldmath$p$})\cdot\hat{\mbox{\boldmath$p$}}\rangle \,=\ 0
\label{06}
\end{equation} 
and
\begin{equation} 
\frac{d~}{dt}\langle\gamma^{\5}\rangle \,=\, i\langle\left[ \mathit{H} , \gamma^{\5}\right]\rangle\,=\,2 \,i\,m \langle\gamma^{\0}\gamma^{\5}\rangle
\label{07}.
\end{equation} 
from which we confirm that the chiral operator $\gamma^{\5}$ is {\em not} a constant of the motion \cite{DeL98}.
The effective value of Eq.~(\ref{07}) appears only when both positive and negative frequencies are taken into account
to compose a Dirac wave-packet, i. e. 
the non-null expectation value of $\langle\gamma_{\0}\gamma_{\5}\rangle$ is revealed by the interference
between Dirac equation solutions with opposite sign frequencies.
The effective contribution due to this interference effect lead us to report to the Dirac wave-packet formalism
in order to quantify neutrino chiral oscillations in the presence of an external magnetic field.
 
Assuming the simplifying hypothesis of a uniform magnetic field  $\mbox{\boldmath$B$}$,
the physical implications of the non-minimal coupling with an external magnetic field can then be studied by means of the
eigenvalue problem expressed by the Hamiltonian equation
\begin{eqnarray} 
H\bb{\mbox{\boldmath$p$}} \, \varphi_{\n} = \,E_{\n}\bb{\mbox{\boldmath$p$}} \, \varphi_{\n}
 		   &=& \left\{\mbox{\boldmath$\alpha$}\cdot \mbox{\boldmath$p$} + \beta \left[m - \mu\, \mbox{\boldmath$\Sigma$}\cdot \mbox{\boldmath$B$}\right]\right\}\varphi_{\n}
\label{10},~~
\end{eqnarray} 
for which the explicit $4 \times 4$ matrix representation is given by
\begin{equation} 
\fl H\bb{\mbox{\boldmath$p$}} \varphi_{\n} = \left[\begin{array}{cccc}
m - \mu B_{\z} & -\mu (B_{\x} - i B_{\y})& p_{\z} & p_{\x} - i p_{\y}\\
-\mu (B_{\x} + i B_{\y})& m + \mu B_{\z} & p_{\x} + i p_{\y} & -p_{\z}\\
p_{\z} & p_{\x} - i p_{\y}& - (m - \mu B_{\z}) & \mu (B_{\x} - i B_{\y})\\
 p_{\x} + i p_{\y} & -p_{\z} & \mu (B_{\x} + i B_{\y})& -(m + \mu B_{\z}) 
\end{array}\right] \varphi_{\n}
\label{10a}.~~
\end{equation}
The most general eigenvalue ($E_{\n}\bb{\mbox{\boldmath$p$}}$) solution of the above problem is given by 
\begin{eqnarray} 
\fl E_{\n}\bb{\mbox{\boldmath$p$}} = \,  \pm E_{\s}\bb{\mbox{\boldmath$p$}} 
&=& \pm \sqrt{m^{\2} + \mbox{\boldmath$p$}^{\2} + \mbox{\boldmath$a$}^{\2} +\bb{\mi 1}^{\s}2
\sqrt{m^{\2}\mbox{\boldmath$a$}^{\2} + \bb{\mbox{\boldmath$p$} \times \mbox{\boldmath$a$}}^{\2}}}, ~~~~s\,=\, 1,\,2
\label{11},
\end{eqnarray} 
where we have denoted $E_{\n \ig \1,\2,\3,\4} = \pm E_{\s \ig \1,\2}$
and we have set $\mbox{\boldmath$a$} = \mu\, \mbox{\boldmath$B$}$.
The complete set of orthonormal eigenstates $\varphi_{\n}$ thus can be written in terms of the eigenfunctions $\mathcal{U}\bb{p_{\s}}$ with positive energy eigenvalues ($+ E_{\s}\bb{\mbox{\boldmath$p$}}$)
and the eigenfunctions $\mathcal{V}\bb{p_{\s}}$ with negative energy eigenvalues ($- E_{\s}\bb{\mbox{\boldmath$p$}}$),
\begin{eqnarray} 
\mathcal{U}\bb{p_{\s}} &=& -N\bb{p_{\s}}\, \left\{\sqrt{\frac{A^{\mi}_{\s}}{A^{\pl}_{\s}}},\,\sqrt{\frac{\alpha^{\pl}_{\s}}{\alpha^{\mi}_{\s}}},\,\sqrt{\frac{A^{\mi}_{\s}\alpha^{\pl}_{\s}}{A^{\pl}_{\s}\alpha^{\mi}_{\s}}},\,-1\right\}^{\dagger}\nonumber\\
\mathcal{V}\bb{p_{\s}} &=& -N\bb{p_{\s}}\, \left\{\sqrt{\frac{A^{\mi}_{\s}}{A^{\pl}_{\s}}},\,-\sqrt{\frac{\alpha^{\mi}_{\s}}{\alpha^{\pl}_{\s}}},\,-\sqrt{\frac{A^{\mi}_{\s}\alpha^{\mi}_{\s}}{A^{\pl}_{\s}\alpha^{\pl}_{\s}}},\,-1\right\}^{\dagger}
\label{12},
\end{eqnarray}
where $p_{\s}$ is the relativistic {\em quadrimomentum}, $p_{\s} = (E_{\s}\bb{\mbox{\boldmath$p$}}, \mbox{\boldmath$p$})$,
$N\bb{p_{\s}}$ is the normalization constant and
\begin{equation} 
A^{\ppm}_{\s} = \Delta_{\s}^{\2}\bb{\mbox{\boldmath$p$}} \pm 2 m |\mbox{\boldmath$a$}|  - \mbox{\boldmath$a$}^{\2},~~~
\alpha^{\ppm}_{\s}= 2 E_{\s}\bb{\mbox{\boldmath$p$}} |\mbox{\boldmath$a$}| \pm (\Delta_{\s}^{\2}\bb{\mbox{\boldmath$p$}} + \mbox{\boldmath$a$}^{\2})
\nonumber
\end{equation}
with
\begin{equation}
\Delta_{\s}^{\2}\bb{\mbox{\boldmath$p$}} = E_{\s}^{\2}\bb{\mbox{\boldmath$p$}} - (m^{\2} + \mbox{\boldmath$p$}^{\2}) + \mbox{\boldmath$a$}^{\2}.
\end{equation}
We can observe that the above spinorial solutions are free of any additional constraint,
namely, at a given time $t$, they are independent functions of $\mbox{\boldmath$p$}$ and they do not represent chirality/helicity eigenstates.

In order to describe the above Hamiltonian dynamics for a generic observable $\mathcal{O}\bb{t}$
we can firstly seek a generic plane-wave decomposition as
\begin{eqnarray}
&& \exp{[- i(E_{\s}\bb{\mbox{\boldmath$p$}}\,t -\mbox{\boldmath$p$} \cdot \mbox{\boldmath$x$})]}\,\mathcal{U}\bb{p_{\s}},
~~~~\mbox{for positive frequencies and}\nonumber\\
&& \exp{[~ i(E_{\s}\bb{\mbox{\boldmath$p$}}\,t -\mbox{\boldmath$p$} \cdot \mbox{\boldmath$x$})]}\,\mathcal{V}\bb{p_{\s}},
 ~~~~\mbox{for negative frequencies},
\label{13}
\end{eqnarray}
so that the time-evolution of a plane-wave-packet $\psi\bb{t, \mbox{\boldmath$x$}}$ can be written as
\begin{eqnarray}
\psi\bb{t, \mbox{\boldmath$x$}}
&=& \int\hspace{-0.1 cm} \frac{d^{\3}\hspace{-0.1cm}\mbox{\boldmath$p$}}{(2\pi)^{\3}}
\sum_{\s \ig \1,\2}\{b\bb{p_{\s}}\mathcal{U}\bb{p_{\s}}\, \exp{[- i\,E_{\s}\bb{\mbox{\boldmath$p$}}\,t]}
\nonumber\\
&&~~~~~~~~~~~~~~~~
+ d^*\bb{\tilde{p}_{\s}}\mathcal{V}\bb{\tilde{p}_{\s}}\, \exp{[+i\,E_{\s}\bb{\mbox{\boldmath$p$}}\,t]}\}
\exp{[i \, \mbox{\boldmath$p$} \cdot \mbox{\boldmath$x$}]},
\label{14}
\end{eqnarray}
with $\tilde{p}_{\s} = (E_{\s},-\mbox{\boldmath$p$})$.
Meanwhile, the Eq.~(\ref{14}) requires some extensive mathematical manipulations
for explicitly constructing the dynamics of an operator $\mathcal{O}\bb{t}$ like
\begin{equation}
\mathcal{O}\bb{t} = \int{d^{\3}\mbox{\boldmath$x$}\,\psi^{\dagger}\bb{t, \mbox{\boldmath$x$}}\,\mathcal{O}\,
\psi\bb{t, \mbox{\boldmath$x$}}}
\label{15}.
\end{equation}
If, however, the quoted observables like the chirality $\gamma^{\5}$,
the helicity $h$ or even the spin projection onto $\mbox{\boldmath$B$}$
commuted with the Hamiltonian $H$, we could reconfigure the above solutions to simpler ones. 
To illustrate this point, let us limit our analysis to very restrictive spatial configurations of $\mbox{\boldmath$B$}$ so that,
as a first attempt, we can calculate the observable expectation values which appear in Eq.~(\ref{04}).
Let us then assume that the magnetic field $\mbox{\boldmath$B$}$ is either orthogonal or parallel to the momentum $\mbox{\boldmath$p$}$.
For both of these cases the spinor eigenstates can then be decomposed into orthonormal bi-spinors as
\begin{equation}
\mathcal{U}\bb{p_{\s}} = N^{\pl}\bb{p_{\s}}\left[\begin{array}{r} \varphi^{\pl}\bb{p_{\s}}\\ \chi^{\pl}\bb{p_{\s}}\end{array}\right]
\end{equation}
and
\begin{equation} 
\mathcal{V}\bb{p_{\s}} = N^{\mi}\bb{p_{\s}}\left[\begin{array}{r} \varphi^{\mi}\bb{p_{\s}}\\ \chi^{\mi}\bb{p_{\s}}\end{array}\right]
\label{16}.
\end{equation}
Eventually, in order to simplify some subsequent calculations involving chiral oscillations, we could set 
$\varphi^{\ppm}_{\1,\2}$ and $\chi^{\ppm}_{\1,\2}$
as eigenstates of the spin projection operator $\mbox{\boldmath$\sigma$}\cdot\mbox{\boldmath$B$}$,
i. e. beside of being energy eigenstates, the general solutions $\mathcal{U}\bb{p_{\s}}$
and $\mathcal{V}\bb{p_{\s}}$ would become eigenstates of the operator $\mbox{\boldmath$\Sigma$}\cdot\mbox{\boldmath$B$}$ and, equivalently, of $\mbox{\boldmath$\Sigma$}\cdot\mbox{\boldmath$a$}$.

Now the Eq.(\ref{10}) can be decomposed into a pair of coupled equations like
\begin{eqnarray}
\left(\pm E_{\s} - m + \mbox{\boldmath$\sigma$}\cdot\mbox{\boldmath$a$} \right)\varphi^{\ppm}_{\s} &=& \pm \mbox{\boldmath$\sigma$}\cdot\mbox{\boldmath$p$}\,\chi^{\ppm}_{\s},\nonumber\\
\left(\pm E_{\s} + m - \mbox{\boldmath$\sigma$}\cdot\mbox{\boldmath$a$} \right)\chi^{\ppm}_{\s} &=& \pm \mbox{\boldmath$\sigma$}\cdot\mbox{\boldmath$p$}\,\varphi^{\ppm}_{\s},
\label{17}
\end{eqnarray}
where we have suppressed the $p_{\s}$ dependence.
By introducing the commuting relation 
$[\mbox{\boldmath$\sigma$}\cdot\mbox{\boldmath$p$},\,\mbox{\boldmath$\sigma$}\cdot\mbox{\boldmath$B$}] = 0$ 
which is derived when $\mbox{\boldmath$p$}\times\mbox{\boldmath$B$} = 0$,
the eigenspinor representation can be reduced to
\begin{equation}
\mathcal{U}\bb{p_{\s}} = \sqrt{\frac{E_{\s} + m_{\s}}{2E_{\s}}}
\left[\begin{array}{r} \varphi^{\pl}_{\s}\\ \frac{\mbox{\boldmath$\sigma$}\cdot\mbox{\boldmath$p$}}{E_{\s}+ m_{\s}}\,\varphi^{\pl}_{\s}\end{array}\right]\end{equation} and \begin{equation} 
\mathcal{V}\bb{p_{\s}} = \sqrt{\frac{E_{\s} + m_{\s}}{2E_{\s}}}
\left[\begin{array}{r} \frac{\mbox{\boldmath$\sigma$}\cdot\mbox{\boldmath$p$}}{E{\s}+ m_{\s}}\,\chi^{\mi}_{\s}\\ \chi^{\mi}_{\s}\end{array}\right]
\label{25},
\end{equation}
with $m_{\s} = m - \bb{\mi 1}^{\s}|\mbox{\boldmath$a$}|$
and the energy eigenvalues
\begin{equation}
\pm E_{\s} = \pm \sqrt{\mbox{\boldmath$p$}^{\2} + m_{\s}^{\2}}
\label{26}.
\end{equation}
In this case, the closure relations can be constructed in terms of 
\begin{eqnarray}
\sum_{\s \ig \1,\2}{\mathcal{U}\bb{p_{\s}}\otimes\mathcal{U}^{\dagger}\bb{p_{\s}}\gamma_{\0}}&=&
\sum_{\s \ig \1,\2}{\left\{\frac{\gamma_{\mu}p_{\s}^{\mu} + m_{\s}}{2 E_{\s}} 
\left[\frac{1-\bb{\mi 1}^{\s}\mbox{\boldmath$\Sigma$}\cdot\hat{\mbox{\boldmath$a$}}}{2}\right]\right\}}\nonumber\\
-\sum_{\s \ig \1,\2}{\mathcal{V}\bb{p_{\s}}\otimes\mathcal{V}^{\dagger}\bb{p_{\s}}\gamma_{\0}}&=&
 \sum_{\s \ig \1,\2}{\left\{\frac{-\gamma_{\mu}p_{\s}^{\mu} + m_{\s}}{2 E_{\s}} \left[\frac{1-\bb{\mi 1}^{\s}\mbox{\boldmath$\Sigma$}\cdot\hat{\mbox{\boldmath$a$}}}{2}\right]\right\}}.
\label{27}
\end{eqnarray}
Analogously, by introducing the anti-commuting relation
$\{\mbox{\boldmath$\sigma$}\cdot\mbox{\boldmath$p$},\,\mbox{\boldmath$\sigma$}\cdot\mbox{\boldmath$B$}\}$ when 
$\mbox{\boldmath$p$}\cdot\mbox{\boldmath$B$} = 0$, 
the eigenspinor representation can be reduced to
\begin{equation}
\mathcal{U}\bb{p_{\s}} = \sqrt{\frac{\varepsilon_{\0} + m}{2\varepsilon_{\0}}}
\left[\begin{array}{r} \varphi^{\pl}_{\s}\\ \frac{\mbox{\boldmath$\sigma$}\cdot\mbox{\boldmath$p$}}{\varepsilon_{\0}+ m}\,\varphi^{\pl}_{\s}\end{array}\right]\end{equation} and \begin{equation} 
\mathcal{V}\bb{p_{\s}} = \sqrt{\frac{\varepsilon_{\0} + m}{2\varepsilon_{\0}}}
\left[\begin{array}{r} \frac{\mbox{\boldmath$\sigma$}\cdot\mbox{\boldmath$p$}}{\varepsilon_{\0}+ m}\,\chi^{\mi}_{\s}\\ \chi^{\mi}_{\s}\end{array}\right]
\label{21},
\end{equation}
with $\varepsilon_{\0} = \sqrt{\mbox{\boldmath$p$}^{\2} + m^{\2}}$
and the energy eigenvalues
\begin{equation}
\pm E_{\s} = \pm\left[\varepsilon_{\0} + \bb{\mi 1}^{\s}|\mbox{\boldmath$a$}|\right]
\label{22}.
\end{equation}
In this case, the closure relations can be constructed in terms of 
\begin{eqnarray}
\sum_{\s \ig \1,\2}{\mathcal{U}\bb{p_{\s}}\otimes\mathcal{U}^{\dagger}\bb{p_{\s}}\gamma_{\0}}&=&
\frac{\gamma_{\mu}p_{\0}^{\mu} + m}{2 \varepsilon_{\0}} \sum_{\s \ig \1,\2}
{\left[\frac{1-\bb{\mi 1}^{\s}\gamma_{\0}\mbox{\boldmath$\Sigma$}\cdot\hat{\mbox{\boldmath$a$}}}{2}\right]}\nonumber\\
-\sum_{\s \ig \1,\2}{\mathcal{V}\bb{p_{\s}}\otimes\mathcal{V}^{\dagger}\bb{p_{\s}}\gamma_{\0}}&=&
\frac{-\gamma_{\mu}p_{\0}^{\mu} + m}{2 \varepsilon_{\0}} \sum_{\s \ig \1,\2}
{\left[\frac{1-\bb{\mi 1}^{\s}\gamma_{\0}\mbox{\boldmath$\Sigma$}\cdot\hat{\mbox{\boldmath$a$}}}{2}\right]},
\label{23}
\end{eqnarray}
where $p_{\0} = (\varepsilon_{\0}, \mbox{\boldmath$p$})$.

Since we can set $\varphi^{\pl}_{\1,\2} \equiv \chi^{\mi}_{\1,\2}$
as the components of an orthonormal basis,
we can immediately deduce the orthogonality relations
\begin{eqnarray}
&&\mathcal{U}^{\dagger}\bb{p_{\s}} \, \mathcal{U}\bb{p_{\r}} = 
\mathcal{V}^{\dagger}\bb{p_{\s}} \, \mathcal{V}\bb{p_{\r}} = \delta_{\s\r},
\nonumber\\
&&\mathcal{U}^{\dagger}\bb{p_{\s}} \,\gamma_{\0}\, \mathcal{V}\bb{p_{\r}} = 
\mathcal{V}^{\dagger}\bb{p_{\s}} \,\gamma_{\0}\, \mathcal{U}\bb{p_{\r}} = 0
\label{20}
\end{eqnarray}
which are valid for both of the above cases.

Finally, the calculation of the expectation value of $\gamma_{\5}\bb{t}$ is substantially simplified when
we substitute the above closure relations into the wave-packet expression of Eq.~(\ref{14}).
To clarify this point, we suppose
the initial condition over $\psi\bb{t,\mbox{\boldmath$x$}}$ can be set in terms of the Fourier transform of the weight function 
\begin{equation}
\varphi\bb{\mbox{\boldmath$p$}-\mbox{\boldmath$p$}_{\ii}}\,w \,=\,
\sum_{\s \ig \1,\2}{\{b\bb{p_{\s}}\mathcal{U}\bb{p_{\s}} + d^*\bb{\tilde{p}{\s}}\mathcal{V}\bb{\tilde{p}_{\s}}\}}
\label{28}
\end{equation}
so that
\begin{equation}
\psi\bb{0, \mbox{\boldmath$x$}}
= \int\hspace{-0.1 cm} \frac{d^{\3}\hspace{-0.1cm}\mbox{\boldmath$p$}}{(2\pi)^{\3}}
\varphi\bb{\mbox{\boldmath$p$}-\mbox{\boldmath$p$}_{\ii}}\exp{[i \, \mbox{\boldmath$p$} \cdot \mbox{\boldmath$x$}]}
\,w
\label{29}
\end{equation}
where $w$ is some fixed normalized spinor.
By using the orthonormality properties of Eq.~(\ref{20}),
we find \cite{Zub80}
\begin{eqnarray}
b\bb{p_{\s}} &=& \varphi\bb{\mbox{\boldmath$p$}- \mbox{\boldmath$p$}_{\ii}} \, \mathcal{U}^{\dagger}\bb{p_{\s}} \, w, \nonumber\\
d^*\bb{\tilde{p}_{\s}} &=& \varphi\bb{\mbox{\boldmath$p$}- \mbox{\boldmath$p$}_{\ii}}\,\mathcal{V}^{\dagger}\bb{\tilde{p}_{\s}}\, w.
\label{30}
\end{eqnarray}
For {\em any} initial state $\psi\bb{0, \mbox{\boldmath$x$}}$ given by Eq.~(\ref{29}),
the negative frequency solution coefficient
$d^*\bb{\tilde{p}_{\s}}$ necessarily provides a non-null contribution to the time-evolving wave-packet.
This obliges us to take the complete set of Dirac equation solutions to construct a complete and correct wave-packet solution.
Only if we consider the initial spinor $w$ being a positive energy ($E_{\s}\bb{\mbox{\boldmath$p$}}$) and momentum 
$\mbox{\boldmath$p$}$ eigenstate, the contribution due to the negative frequency solutions 
$d^*\bb{\tilde{p}_{\s}}$ will become null and we will have a simple expression for the time-evolution of
any physical observable.
By substituting the closure relations of Eqs.~(\ref{27}) and (\ref{23}) into the time-evolution expression
for the above wave-packet, the Eq.(\ref{14}) can thus be rewritten as
{\small \begin{eqnarray}
\fl\hspace{-0.1 cm}\psi\bb{t, \mbox{\boldmath$x$}}
&\hspace{-0.1 cm}=&
\hspace{-0.1 cm}\int\hspace{-0.1 cm} \frac{d^{\3}\hspace{-0.1cm}\mbox{\boldmath$p$}}{(2\pi)^{\3}}
\varphi\bb{\mbox{\boldmath$p$}-\mbox{\boldmath$p$}_{\ii}}\exp{[i \, \mbox{\boldmath$p$} \cdot \mbox{\boldmath$x$}]}
\sum_{\s \ig \1,\2}\mbox{$\left\{\left[\cos{[E_{\s}\,t]} -i\frac{H_{\s}}{E_{\s}}\sin{[E_{\s}\,t]}\right]\left(\frac{1-(\mi 1)^{\s}\mbox{\boldmath$\Sigma$}\cdot\hat{\mbox{\boldmath$a$}}}{2}\right)\right\}$}
w~~
\label{14A}
\end{eqnarray}}
for the first case where $E_{\s}$ is given by Eq.~(\ref{26}) and $H_{\s} = \mbox{\boldmath$\alpha$}\cdot \mbox{\boldmath$p$} + \gamma_{\0} m_{\s}$, or as
{\small \begin{eqnarray}
\fl\hspace{-0.1 cm}\psi\bb{t, \mbox{\boldmath$x$}}
&\hspace{-0.1 cm}=&
\hspace{-0.1 cm}\int\hspace{-0.1 cm} \frac{d^{\3}\hspace{-0.1cm}\mbox{\boldmath$p$}}{(2\pi)^{\3}}
\varphi\bb{\mbox{\boldmath$p$}-\mbox{\boldmath$p$}_{\ii}}\exp{[i \, \mbox{\boldmath$p$} \cdot \mbox{\boldmath$x$}]}
\sum_{\s \ig \1,\2}\mbox{$\left\{\left[\cos{[E_{\s}\,t]} -i\frac{H_{\0}}{\varepsilon_{\0}}\sin{[E_{\s}\,t]}\right]\left(\frac{1-(\mi 1)^{\s}\gamma_{\0}\mbox{\boldmath$\Sigma$}\cdot\hat{\mbox{\boldmath$a$}}}{2}\right)\right\}$}
w~~~~
\label{14B}
\end{eqnarray}}
for the second case where $\varepsilon_{\0}$ is given by Eq.~(\ref{22}) and $H_{\0} = \mbox{\boldmath$\alpha$}\cdot \mbox{\boldmath$p$} + \gamma_{\0} m$.

Once we have assumed the neutrino electroweak interactions at the source and detector are ({\em left}) chiral
$\left(\overline{\psi} \gamma^{\mu}(1 - \gamma^{\5})\psi W_{\mu}\right)$
only the component with negative chirality contributes to the propagation. 
It was already demonstrated that, in vacuum, chiral oscillations can introduce very small modifications
to the neutrino conversion formula \cite{DeL98,Ber05}.
The probability of a neutrino produced as a negative chiral eigenstate be detected after a time $t$ can be summarized by the expression
\begin{eqnarray}
P(\mbox{\boldmath$\nu_{\alpha,\L}$}\rightarrow\mbox{\boldmath$\nu_{\alpha,\L}$};t) 
& = &
\int{d^{\3}\mbox{\boldmath$x$}\,\psi^{\dagger}\bb{t, \mbox{\boldmath$x$}}\,\frac{1 - \gamma_{\5}}{2}\,
\psi\bb{t, \mbox{\boldmath$x$}}} = \frac{1}{2}\left(1 - \langle\gamma_{\5}\rangle\bb{t}\right)
\label{15A}.
\end{eqnarray}
From this integral, it is readily seen that an initial $\mi 1$ chiral
mass-eigenstate will evolve with time changing its chirality.
By assuming the fermionic particle is created  at time $t=0$ as a $\mi 1$ chiral eigenstate ($\gamma_{\5} w = \mi w$),
in case of $[\mbox{\boldmath$\sigma$}\cdot\mbox{\boldmath$p$},\,\mbox{\boldmath$\sigma$}\cdot\mbox{\boldmath$B$}] = 0$
({\boldmath$B$} parallel to {\boldmath$p$}), we could write
{\small \begin{eqnarray}
\fl \langle\gamma_{\5}\rangle\bb{t}&=&  
\int\hspace{-0.1 cm} \frac{d^{\3}\hspace{-0.1cm}\mbox{\boldmath$p$}}{(2\pi)^{\3}}\varphi^{\2}\bb{\mbox{\boldmath$p$}-\mbox{\boldmath$p$}_{\ii}}\times\nonumber\\
\fl&&~~
w^{\dagger}\sum_{\s \ig \1,\2}\mbox{$\left\{\left[\gamma_{\5}\cos^{\2}{[E_{\s}\,t]} +i \frac{[H_{\s},\,\gamma_{\5}]}{2E_{\s}}\sin{[2\,E_{\s}\,t]} + \frac{H_{\s}\,\gamma_{\5}\,H_{\s}}{E_{\s}^{\2}}\sin^{\2}{[E_{\s}\,t]}\right]\left(\frac{1-(\mi 1)^{\s}\mbox{\boldmath$\Sigma$}\cdot\hat{\mbox{\boldmath$a$}}}{2}\right)\right\}$}\,w\nonumber\\
\fl&=&(\mi 1) \int\hspace{-0.1 cm} \frac{d^{\3}\hspace{-0.1cm}\mbox{\boldmath$p$}}{(2\pi)^{\3}}\varphi^{\2}\bb{\mbox{\boldmath$p$}-\mbox{\boldmath$p$}_{\ii}}
\sum_{\s \ig \1,\2}\mbox{$\left\{\left[\cos^{\2}{[E_{\s}\,t]} + \frac{\mbox{\boldmath$p$}^{\2} - m_{\s}^{\2}}{E_{\s}^{\2}}\sin^{\2}{[E_{\s}\,t]}\right]\,w^{\dagger}\left(\frac{1-(\mi 1)^{\s}\mbox{\boldmath$\Sigma$}\cdot\hat{\mbox{\boldmath$a$}}}{2}\right)w\right\}$}\nonumber\\
\fl&=&(\mi 1) \int\hspace{-0.1 cm} \frac{d^{\3}\hspace{-0.1cm}\mbox{\boldmath$p$}}{(2\pi)^{\3}}\varphi^{\2}\bb{\mbox{\boldmath$p$}-\mbox{\boldmath$p$}_{\ii}}
\sum_{\s \ig \1,\2}\mbox{$\left\{\left[\frac{\mbox{\boldmath$p$}^{\2}}{E_{\s}^{\2}} + \frac{m_{\s}^{\2}}{E_{\s}^{\2}}\cos{[2\,E_{\s}\,t]}\right]\,w^{\dagger}\left(\frac{1-(\mi 1)^{\s}\mbox{\boldmath$\Sigma$}\cdot\hat{\mbox{\boldmath$a$}}}{2}\right)w\right\}$}
\label{15B},
\end{eqnarray}}
where we have used the wave packet expression of Eq.~(\ref{14A}) and, in the second passage, we have observed that
\begin{equation}
w^{\dagger} \gamma_{\5} w = \mi 1,~~~~  w^{\dagger} [H_{\s},\,\gamma_{\5}]w = 0 ~~~~\mbox{and}~~~~ H_{\s}\,\gamma_{\5}\,H_{\s} = \mbox{\boldmath$p$}^{\2} - m_{\s}^{\2}.
\label{15B1}
\end{equation}
The above expression can be reduced to a simpler one in the non-interacting case \cite{DeL98}.
Due to a residual interaction with the external magnetic field {\boldmath$B$} we could also observe chiral oscillations
in the ultra-relativistic limit.
However, from the phenomenological point of view, the coefficient of the oscillating term goes with $\frac{m_{\s}^{\2}}{E_{\s}^{\2}}$ which makes
chiral oscillations become not relevant for ultra-relativistic neutrinos \cite{Ber05,Ber05A}.
As a {\em toy model} illustration, by assuming a highly peaked momentum distribution centered around
a non-relativistic momentum $p_{\ii}\ll m_{\s}$, where the wave packet effects are practically ignored, the chiral conversion formula
can be written as
\begin{equation}
\fl P(\mbox{\boldmath$\nu_{\alpha,\L}$}\rightarrow\mbox{\boldmath$\nu_{\alpha,\R}$};t) 
 \approx  \frac{1}{2}\left(1 - \cos{[2\,m\,t]}\cos{[2\,|\mbox{\boldmath$a$}|\,t]}
 -\sin {[2\,m\,t]}\sin{[2\,|\mbox{\boldmath$a$}|\,t]}w^{\dagger}\mbox{\boldmath$\Sigma$}\cdot\hat{\mbox{\boldmath$a$}}w\right)
\label{15A222}
\end{equation}
where all the oscillating terms come from the interference between positive and negative frequency solutions which compose the wave packets.
Turning back to the case where
$\{\mbox{\boldmath$\sigma$}\cdot\mbox{\boldmath$p$},\,\mbox{\boldmath$\sigma$}\cdot\mbox{\boldmath$B$}\} = 0$
({\boldmath$B$} orthogonal to {\boldmath$p$}), we could have a phenomenologically more interesting result. 
By following a similar procedure with the mathematical manipulations, we could write
\begin{eqnarray}
\fl \langle\gamma_{\5}\rangle\bb{t} &=&  
\int\hspace{-0.1 cm} \frac{d^{\3}\hspace{-0.1cm}\mbox{\boldmath$p$}}{(2\pi)^{\3}}\varphi^{\2}\bb{\mbox{\boldmath$p$}-\mbox{\boldmath$p$}_{\ii}}
w^{\dagger}\mbox{$\left\{\gamma_{\5}\cos{[E_{\1}\,t]}\cos{[E_{\2}\,t]} + \frac{H_{\0}\,\gamma_{\5}\,H_{\0}}{\epsilon_{\0}^{\2}} \sin{[E_{\1}\,t]}\sin{[E_{\2}\,t]} +\right.$}\nonumber\\
\fl&&~~~~~~~~~~~~~~~~~~~~~~~~~\left.\mbox{$\frac{i}{2}\left[[H_{\0},\,\gamma_{\5}]\sin{[E_{\1}+E_{\2}]} - \{H_{\0},\,\gamma_{\5}\}\gamma_{\0}\mbox{\boldmath$\Sigma$}\cdot\hat{\mbox{\boldmath$a$}}\sin{[E_{\1}-E_{\2}]} \right]$}\right\}w\nonumber\\
\fl&=&(\mi 1) \int\hspace{-0.1 cm} \frac{d^{\3}\hspace{-0.1cm}\mbox{\boldmath$p$}}{(2\pi)^{\3}}\varphi^{\2}\bb{\mbox{\boldmath$p$}-\mbox{\boldmath$p$}_{\ii}}
\mbox{$\left\{\cos{[E_{\1}\,t]}\cos{[E_{\2}\,t]} + \frac{\mbox{\boldmath$p$}^{\2} - m^{\2}}{\epsilon_{\0}^{\2}} \sin{[E_{\1}\,t]}\sin{[E_{\2}\,t]}\right\}$}\nonumber\\
\fl&=&(\mi 1) \int\hspace{-0.1 cm} \frac{d^{\3}\hspace{-0.1cm}\mbox{\boldmath$p$}}{(2\pi)^{\3}}\varphi^{\2}\bb{\mbox{\boldmath$p$}-\mbox{\boldmath$p$}_{\ii}}
\mbox{$\left\{\frac{\mbox{\boldmath$p$}^{\2}}{\epsilon_{\0}^{\2}}\cos{[2\,|\mbox{\boldmath$a$}|\,t]}+ \frac{ m^{\2}}{\epsilon_{\0}^{\2}} \cos{[2\,\epsilon_{\0}\,t]}\right\}$}
\label{15C},
\end{eqnarray}
where we have used the wave packet expression of Eq.~(\ref{14B}) and, in addition to $w^{\dagger} \gamma_{\5} w = \mi 1$,
we have also observed that
$\{H_{\0},\,\gamma_{\5}\} = 2 \gamma_{\5}\mbox{\boldmath$\Sigma$}\cdot\hat{\mbox{\boldmath$p$}}$ and, subsequently, 
$w^{\dagger}\mbox{\boldmath$\Sigma$}\cdot\hat{\mbox{\boldmath$p$}}\gamma_{\0}\mbox{\boldmath$\Sigma$}\cdot\hat{\mbox{\boldmath$a$}} w = 0$.
Now, in addition to the non-interacting oscillating term $\frac{ m^{\2}}{\epsilon_{\0}^{\2}} \cos{[2\,\epsilon_{\0}\,t]}$, which comes from
the interference between positive and negative frequency solutions of the Dirac equation, we have
an extra term which comes from the interference between equal sign frequencies and, for very large time scales, can substantially change the oscillating results. 
In this case, it is interesting to observe that the ultra-relativistic limit of Eq.(\ref{15C}) leads to the following
expressions for the chiral conversion formulas,
\begin{equation}
P(\mbox{\boldmath$\nu_{\alpha,\L}$}\rightarrow\mbox{\boldmath$\nu_{\alpha,\R}$};t) 
 \approx  \frac{1}{2}\left(1 - \cos{[2\,|\mbox{\boldmath$a$}|\,t]}\right)
\label{15A2}
\end{equation}
and
\begin{equation}
P(\mbox{\boldmath$\nu_{\alpha,\L}$}\rightarrow\mbox{\boldmath$\nu_{\alpha,\L}$};t) 
 \approx  \frac{1}{2}\left(1 + \cos{[2\,|\mbox{\boldmath$a$}|\,t]}\right)
\label{15A3}
\end{equation}
which, differently from chiral oscillations in vacuum, can be phenomenologically relevant.
Obviously, we are reproducing the consolidated results already attributed to neutrino spin-flipping
\cite{Kim93} where, by taking the ultra-relativistic limit, the chirality quantum number can be approximated by the
helicity quantum number, but now it was accurately derived from the complete formalism with Dirac spinors.
We still remark that, in the standard treatment of
vacuum neutrino oscillations, the use of scalar mass-eigenstate wave packets made up
exclusively of positive frequency plane-wave solutions is implicitly
assumed. A satisfactory description
of fermionic (spin one-half) particles requires the use of the
Dirac equation as evolution equation for the mass-eigenstates despite the standard oscillation formula giving
the correct result when it is {\em properly} interpreted.
We have observed in Eqs.~(\ref{15C}) and (\ref{15B}) that the spinorial form and
the interference between positive and negative frequency
components of the mass-eigenstate wave packets
can introduce small modifications
to the {\em standard} conversion formulas when they concern with
{\em non-relativistic} neutrinos \cite{Ber05A}.
In the next step, by following the Dirac wave packet prescription, we intend to study the
coupling of chiral oscillations with flavor conversion effects for neutrinos non-minimally
interacting with an external magnetic field and, subsequently, verify the possibility of some phenomenological
implications.

To summarize, we would have been dishonest if we had ignored the complete analysis of the general case comprised by
Eqs.~(\ref{10}-\ref{12}) where we had not yet assumed an arbitrary (simplified) spatial configuration for the magnetic field.
Meanwhile, such a general case leads to the formal connection between quantum oscillation phenomena and
a very different field. It concerns with the curious fact that those complete (general) expressions
for propagating wave packets
do not satisfy the standard dispersion relations like $E^{\2} = m^{\2}+{\mbox{\boldmath$p$}}^{\2}$ excepting by the two
particular cases where $E_{\s}\bb{\mbox{\boldmath$p$}}^{\2} = m_{\s}^{\2}+ {\mbox{\boldmath$p$}}^{\2}$ for $\mbox{\boldmath$p$}\times\mbox{\boldmath$B$} = 0$
or $\epsilon_{\0}^{\2} = m^{\2}+ {\mbox{\boldmath$p$}}^{\2}$ for $\mbox{\boldmath$p$}\cdot\mbox{\boldmath$B$} = 0$.
By principle, it could represent an inconvenient obstacle forbidding  the extension of these restrictive cases to
a general one. However, we believe that it can also represent a starting point in discussing 
dispersion relations which may be incorporated into frameworks encoding the breakdown (or the validity) of Lorentz invariance. 
\subsection*{Acknowledgments}
The author thanks FAPESP (PD 04/13770-0) for Financial Support.

\end{document}